\documentclass[fleqn,twoside]{article}
\usepackage{espcrc2}

\usepackage{graphicx}
\usepackage[figuresright]{rotating}

\def\bc{\begin{center}}
\def\ec{\end{center}}
\newcommand{\be}{\begin{equation}}
\newcommand{\ee}{\end{equation}}

\newcommand{\MSbar}{\overline{\mathrm{MS}}}

\newcommand{\nn}{\nonumber}

\newcommand{\as}{\alpha_{ s}}

\newcommand{\bb}{B^{0}-\bar{B}^{0}}
\newcommand{\kk}{K^{0}-\bar{K}^{0}}

\newcommand{\beq}{\begin{equation}}
\newcommand{\eeq}{\end{equation}}
\newcommand{\beqs}{\begin{equation*}}
\newcommand{\eeqs}{\end{equation*}}
\newcommand{\beqn}{\begin{eqnarray}}
\newcommand{\eeqn}{\end{eqnarray}}
\newcommand{\bea}{\begin{eqnarray}}
\newcommand{\eea}{\end{eqnarray}}
\newcommand{\beqns}{\begin{eqnarray*}}
\newcommand{\eeqns}{\end{eqnarray*}}
\newcommand{\mfrac}[2]{\frac{\textstyle #1}{\textstyle #2}}

\def\vdir{v\kern-5.75pt\raise0.15ex\hbox{${\scriptstyle /}$}}
\def\pdir{p\kern-7.8pt\raise0.2ex\hbox{\Big{/}}}
\def\ddir{D\kern-7pt\raise0.2ex\hbox{\big{/}}}
\def\partdir{\partial\kern-7.6pt\raise0.25ex\hbox{/}}
\def\ddirp{D_{\kern-2.75pt\perp}\kern-11pt\raise0.2ex\hbox{\big{/}}\kern+4.5pt}

\newcommand{\AmS}{{\protect\the\textfont2
  A\kern-.1667em\lower.5ex\hbox{M}\kern-.125emS}}

\title{Non-perturbative renormalisation
of four fermion operators and\\ $\bb$ mixing with Wilson fermions
\thanks{Presented by J.~Reyes at ``Lattice 2002", Boston.}
\thanks{This work has been  supported in part by the EU IHP under HPRN-CT-2000-00145
Hadrons/LatticeQCD.}}

\author{$\mathrm{SPQcdR}$ Collaboration\\
D. Be\'cirevi\'c\address[roma1]{Dip. di Fisica, Univ. di Roma ``La Sapienza''
and INFN-Sezione di Roma, Piazzale A. Moro 2, I-00185 Roma, Italy.},
        V. Gim\'enez \address{Dep. de F\'\i sica Te\`orica and IFIC, Univ. de
	Val\`encia, Dr. Moliner 50, E-46100, Burjassot, Val\`encia, Spain.},
        V. Lubicz\address{Dip. di Fisica, Univ. di Roma Tre and INFN-Sezione di
	Roma III, Via della Vasca
	Navale 84, I-00146 Roma, Italy.},
        G. Martinelli\addressmark[roma1],
	M. Papinutto\address{
      DESY, Theory Group,
      Notkestrasse 85,
      D-22603 Hamburg,
      Germany.},
	J. Reyes\addressmark[roma1]}
       
\begin{document}

\begin{abstract}
We present new results for the renormalisation and
subtraction constants for the four fermion $\Delta F=2$
operators, computed non-perturbatively in the RI-MOM scheme
(in the Landau gauge). From our preliminary analysis of the
lattice data at $\beta=6.45$, for the $\bb$ mixing bag-parameter
we obtain $B_B^{RGI} = 1.46(7)(1)$.

\vspace{1pc}
\end{abstract}

\maketitle

Matrix elements of the four fermion (4f) operators play an important r\^ole in the studies
of CP-violation in the Standard Model and beyond. Of particular interest 
are the matrix elements describing the $\bb$ and $\kk$ mixing
amplitudes. Lattice QCD is at present the best suited method to compute
such quantities non-perturbatively (NP). To keep the theoretical uncertainties 
under control, it is of vital interest to renormalise the corresponding 4f-operators 
non-perturbatively. An additional difficulty arises when working with 
Wilson fermions on the lattice, namely all the parity even ($\Delta F=2$) operators 
mix among themselves. Therefore, besides the overall renormalisation, the spurious
mixing should be subtracted.

\section{Non-perturbative renormalisation of four fermion operators}
To compute the values of the subtraction and renormalisation constants we use
the method described in ref.~\cite{bibbia}. Following those papers, we compute the 
Green functions (GF) of all $\Delta F=2$ operators, sandwiched
by the quark fields, in a specific gauge (in practice we opt for the Landau gauge). 
After projecting the amputated GF's onto all independent Dirac structures we get
$\Gamma_{ij}(p^2)$, on which the following (RI-MOM) renormalisation condition is imposed
\beqn\label{RI-con}
&&\Gamma_{ij}(p^2)|_{\mu^2=-p^2}=\Gamma^{(0)}_{ij}(p^2)\:,
\eeqn
where $\Gamma_{ij}^{(0)}$ denote the tree level values of the GF's. We employed
this prescription in ref.~\cite{Rc-our} to compute the renormalisation and subtraction constants
for all four fermion operators.
Here, we focus
on the parity even ones, {\it i.e.},
\beqn
O_1&=&\bar b\gamma^\mu q\,\bar b \gamma_\mu q+
\bar b\gamma^\mu\gamma_5q\,\bar b \gamma_\mu\gamma_5q\nn\\
O_2&=&\bar b\gamma^\mu q\,\bar b \gamma_\mu q-
\bar b\gamma^\mu\gamma_5q\,\bar b \gamma_\mu\gamma_5q\nn\\
O_3&=&\bar b q\,\bar b  q-
\bar b\gamma_5q\,\bar b \gamma_5q\nn\\
O_4&=&\bar b q\,\bar b  q+
\bar b\gamma_5q\,\bar b \gamma_5q\nn\\
O_5&=&\mfrac{1}{2}\bar b \sigma^{\mu\nu}q\,\bar b
\sigma_{\mu\nu}q\:.
\label{full-op}
\eeqn
As mentioned above, the lattice regularisation with Wilson fermions breaks
chirality,
thus inducing the spurious mixing among all of the above operators.
Therefore to renormalise the operator $O_1$, we should subtract the effects of
this mixing, namely
\be
\hat O_1(\mu)\!=\! Z(\mu\, a) \left[O_1(a)+\sum_{i=2}^5 \Delta_{i}(a)O_i(a)\right]  \: .
\ee
The condition~(\ref{RI-con}) allows us to compute all subtraction
$\Delta_{2-5}(a)$ and the 
renormalisation constant $Z(\mu\, a)$. We perform such a calculation for $4$
different values of  the lattice spacing, corresponding to $\beta=6.0$, $6.2$,
$6.4$, $6.45$.  At each $\beta$, we work with $4$ values of the light quark mass,
which allows us to extrapolate  to the chiral limit (in which the RI-MOM
scheme is defined). In this extrapolation a special attention is
given to the subtractions of the  
pseudo-Goldstone 
boson (PGB) pole. These artefacts are the consequence of the fact that  the
operators in our Green
functions are inserted at zero momentum. To subtract the effects of
the PGB pole, we extend the method of ref.~\cite{pgb1}, in a way described in
ref.~\cite{Bp-our}. 
\begin{figure}[t!]
\rotatebox{-90}{\includegraphics[bb=3cm 1.4cm 17cm 10cm,scale=0.31]{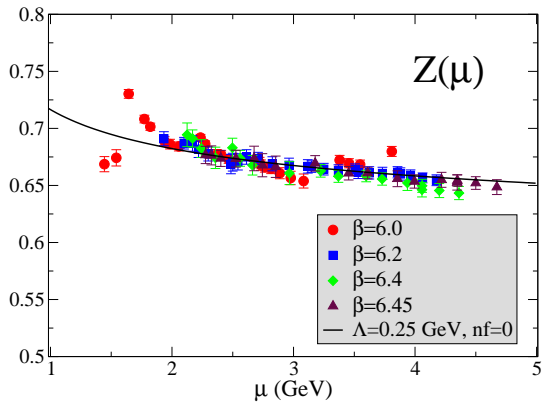}}
\caption{\label{z1:all}\small $Z(\mu)$ as computed at four values of $\beta$ and rescaled to $\beta = 6.45$ as indicated 
in eq.~(\ref{rescale}). The full line correspond to the perturbative running of the operator $O_1(\mu)$ 
at NLO in the RI-MOM scheme (see eq.~(\ref{NLO}) in which we use $\Lambda_{n_f=0}^{\MSbar}=250$~MeV).}
\mbox{}\\[-1cm]
\end{figure}

\begin{figure}
\rotatebox{-90}{\includegraphics[bb=3cm 1.4cm 17cm 10cm,scale=0.31]{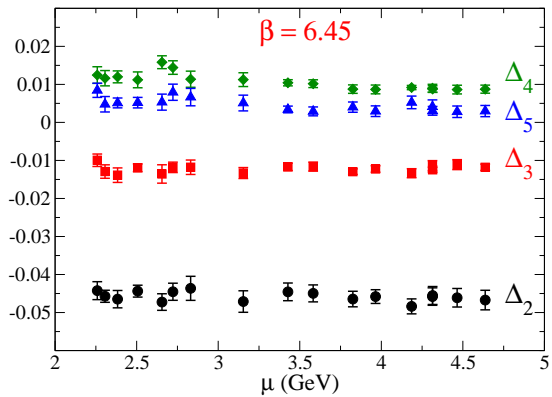}}
\caption{\label{Delta:all}\small Subtraction constants $\Delta_{2-5}$ computed at $\beta = 6.45$ for several values of the 
renormalisation scale $\mu$. We observe that, to a good accuracy, $\partial \Delta_{2-5}/\partial \mu \simeq 0$, 
as it should.}
\end{figure}

In fig.~\ref{z1:all} we plot $Z(\mu\,a)$, obtained for all of our values of $\beta$ and then 
rescaled  to  $\beta=6.45$ according to,
\beqn \label{rescale}
&& Z(\mu\, a')=  R(a',a)\,Z(\mu\, a)\,,
\eeqn
where $R(a',a)$ is $\mu$-independent if the discretisation effects of 
${\cal O}(\mu\, a)$ are negligible. In fig.~\ref{z1:all} we also show  
the renormalisation group running at NLO in perturbation theory, {\it i.e.},
\beqn \label{NLO}
Z(\mu)&=&Z(\mu_0)\,\left({\as(\mu)\over\as(\mu_0)}\right)^{\mfrac{\gamma_0}{2\beta_0}}\times\nn\\
&&\left(1+{\as(\mu_0)-\as(\mu)\over 4
\pi} J_{RI}\right)\:,
\eeqn
where $\beta^{(n_f=0)}_0=11$, $\gamma_0=4$ and
$J^{(n_f=0)}_{RI}=8\log{2}-1933/726$~\cite{bible}. We have chosen $\mu_0\approx
3$~GeV. We see that eq.~(\ref{NLO}) describes our
data quite well. Moreover,  
the ${\cal O}(\mu\, a)$ effects appear to be small, a feature already observed in
the case of the renormalisation constants of the bilinear quark operators,
computed with this same method~\cite{vitt-proc}. 
In fig.~\ref{Delta:all} we show $\Delta_{2-5}$, computed at
$\beta=6.45$. We note that they are very weakly dependent on the renormalisation
scale $\mu$, as one expects in presence of small lattice artefacts. 
We also mention  that their values
decrease with  
the lattice spacing $a$ ({\it i.e.} with larger value of $\beta$).
After fitting each one of them to a constant, we obtain 
\[
\Delta_{2-5}\!=\!\{-4.6(2),-1.2(1),1.0(2),0.4(1)\}\!\!\times 10^{-2}\!\!\,.
\]
For a more detailed discussion, including complete list of numerical results,  
see ref.~\cite{Rc-our}.

\section{$\bb$ mixing parameter}
We now use the results of the previous section 
to compute $B_{1}^{RI}(\mu)$ which parametrises the matrix element of the 
operator $\hat O_1(\mu)$ as
\be
\langle \bar B_q |O_1(\mu)|B_q\rangle= {8\over 3}\,
m_{B_q}^2\,f_{B_q}^2\,{ B_{1}(\mu)}\:.
\ee
Its value is extracted from the plateau of the ratio\\
\scalebox{0.96}{\parbox{5cm}{\[
\frac{ \langle \displaystyle{\sum_{\vec x,\vec y}} P^\dagger (\vec x, -t_x)
  \hat O_1(\vec 0, 0;\mu)  P^\dagger (\vec y, t_y) \rangle}{
  {\displaystyle {8\over 3}}\langle  \displaystyle{\sum_{\vec x} 
A_0 (x) P^\dagger (0)}\rangle \langle  \displaystyle{\sum_{\vec y} 
A_0 (y) P^\dagger (0)}\rangle  }
\!\!\stackrel{t_x,t_y\gg 0}{\longrightarrow}\!\! B_{1}(\mu)\,,
\]}},\\
where $A_\mu$ and $P$ are the heavy-light axial current and the pseudoscalar
density, respectively. Our sample contains $100$ independent gauge field
configuration generated on a $32^3\times70$ lattice in the quenched approximation. 
We work with $4$ light and $6$ heavy quark masses. The directly accessed heavy-light 
pseudoscalar mesons (with light quarks  around the strange quark mass) are
in the range $m_P\in[1.7,3.6]$ GeV. In the above ratio we fix $t_x=21$ and find
the plateau in the interval $t_y\in [45,54]$, for all combinations of our light and
heavy quark masses. We compute $B_1(\mu)$ for $17$ values of the renormalisation
scale $\mu$, which we then convert to the renormalisation group invariant form
(RGI) at NLO accuracy, as 
\be \label{rgi}
B_1^{RGI}\!\!\!=\as(\mu)^{\displaystyle{-{\gamma_0\over 2\beta_0}}}\!\left(\!1+{\as(\mu)\over 4
\pi}J_{RI}\!\right) B^{RI}_1(\mu)\,,
\ee
where $\beta_0$, $\gamma_0$ and $J_{RI}$ are already defined after eq.~(\ref{NLO}).
\begin{figure}
\includegraphics[bb=3.2cm 3cm 10cm 14cm,scale=0.49]{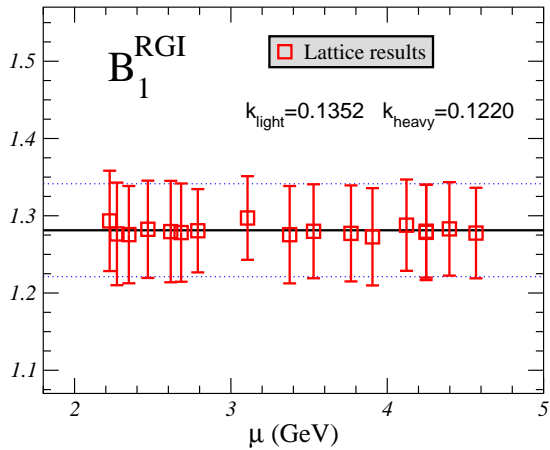}
\caption{\label{BRGI}\small Verification of the scale independence of the parameter $B_1^{RGI}$ 
defined in eq.~(\ref{rgi}).}
\mbox{}\\[-1.2cm]
\end{figure}
In fig.~\ref{BRGI} we show the quality of our $B_1^{RGI}$.

\begin{figure}
\includegraphics[bb=2.5cm 1.2cm 18cm 12cm,angle=-90,scale=0.3]{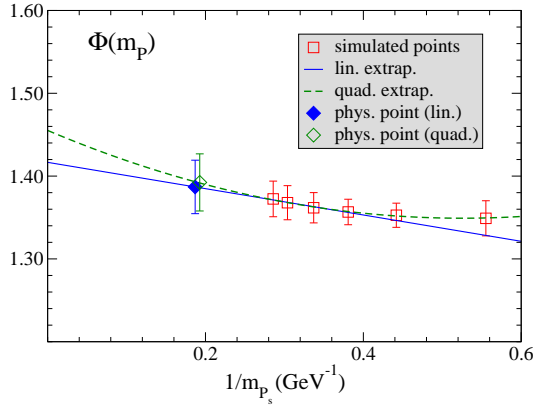}
\caption{\label{H_extr}\small Heavy quark (meson) extrapolation to $1/m_{B_s}$. The linear extrapolation
is made with $4$ heavier mesons only, whereas for the quadratic one all $6$ masses are used.}
\mbox{}\\[-1.1cm]
\end{figure}

In order to get the value of the $B$-parameter relevant for the $\bb$ mixing amplitude, 
we need to extrapolate to the physical point corresponding to $1/m_{B_{d,s}}$.
For that purpose we construct the quantity $\Phi(m_P)$:
\beqn
 \Phi(m_P)\!\!\!\!\!\!&=&\!\!\!\!\!\!\left(\as(m_P)\over
\as(m_B)\right)^{\displaystyle{\gamma_0\over 
2\beta_0}}B_1^{RGI}
\nn\\
&=&\alpha_0+{\alpha_1\over m_P}+{\alpha_2\over m_P^2}+\cdots\, ,
\label{PHI_extr}
\eeqn
where the argument in $\Phi(m_P)$ indicates that the $B$-parameter is computed for
the pseudoscalar meson of mass $m_P$. The r\^ole of the evolution factor in~(\ref{PHI_extr}) 
is to resum the terms $\propto \log{(m_B/m_P)}$, thus removing them from the 
expansion in $1/m_P$. The extrapolated point, $\Phi(m_B)$, is simply
$B_1^{RGI}$, the one that is needed for the $\bb$ mixing. 
The $1/m_P$-extrapolation is illustrated in fig.~\ref{H_extr}. Our preliminary
results read,
\[
B^{RGI}_{B_d}=1.46(7)(1)\,,\quad B^{RGI}_{B_s}=1.38(3)(1)\,,  
\]
where the first error is statistical and the second is the difference
between the results of  the linear and quadratic extrapolations.

\end{document}